\documentclass[12pt]{article}
\usepackage[margin=1in]{geometry}
\usepackage{times}
\usepackage{color}
\usepackage{graphicx}
\usepackage[colorlinks,citecolor=blue,urlcolor=blue,linkcolor=blue,bookmarks=false,hypertexnames=true]{hyperref} 

\usepackage{multirow}
\usepackage{amsmath}
\usepackage{amsthm}
\usepackage{amssymb}
\usepackage{natbib}
\usepackage{rotating}
\usepackage{amsfonts}
\usepackage{bbm}
\usepackage{appendix}

\newtheorem{theorem}{Theorem}

\newtheorem{definition}{Definition}
\usepackage{bbm}

\DeclareMathOperator*{\median}{median}
\DeclareMathOperator*{\argmin}{arg\,min}

\usepackage[linesnumbered,ruled,vlined]{algorithm2e}

\begin{document}
\title{LassoRNet: Accurate dim-light melatonin onset time prediction from multiple blood tissue samples}
\author{
	Michael T. Gorczyca\footnote{Corresponding author: mtg62@cornell.edu} \\
	MTG Research Consulting \and Tavish M. McDonald \\ Lawrence Livermore National Laboratory \and Brandon Oppong-Antwi \\
    Social Security Administration
}
\date{\today}

\maketitle
\begin{abstract}
    \noindent Research on chemotherapy, heart surgery, and vaccines has indicated that the risks and benefits of a treatment could vary depending on the time of day it is administered. A challenge with performing studies on timing treatment administration is that the optimal treatment time is different for each patient, as it would be based on a patient's internal clock time (ICT) rather than the 24-hour day-night cycle time. Prediction methods have been developed to determine a patient's ICT based on biomarker measurements, which can be leveraged to personalize treatment time. However, these methods face two limitations. First, these methods are designed to output predictions given biomarker measurements from a single tissue sample, when multiple tissue samples can be collected over time. Second, these methods are based on linear modelling frameworks, which would not capture the potentially complex relationships between biomarkers and a patient’s ICT. To address these two limitations, this paper introduces a recurrent neural network framework, which we refer to as LassoRNet, for predicting the ICT at which a patient's biomarkers are measured as well as the underlying offset between a patient's ICT and the 24-hour day-night cycle time, or that patient's dim-light melatonin onset (DLMO) time. A novel feature of LassoRNet is a proposed variable selection scheme that minimizes the number of biomarkers needed to predict ICT. We evaluate LassoRNet on three longitudinal circadian transcriptome study data sets where DLMO time was determined for each study participant, and find that it consistently outperforms state-of-the art in both ICT and DLMO time prediction. Notably, LassoRNet obtains a median absolute error of approximately one hour in ICT prediction and 30 to 40 minutes in DLMO time prediction, where DLMO time prediction is performed using three samples collected at sequential time points.
\end{abstract}


\section{Introduction} \label{sec1}

The circadian clock is an internal timekeeping mechanism that coordinates behavioral and physiological processes over the course of a 24-hour day \citep{Lane2022}. This coordination results in biological phenomena such as a person's body temperature \citep{Refinetti1992}, cognitive performance \citep{Blatter2007}, and heart rate \citep{Malpas1990} to display oscillatory behavior over a 24-hour period. An understanding of the times at which these phenomena peak and trough has implications in improving patient care and quality of life, as the risks and benefits of treatments such as chemotherapy \citep{Dallmann2016}, open-heart surgery \citep{Montaigne2018}, and vaccines \citep{Long2016} could differ depending on the time of day they are administered.

A current challenge with performing studies on timing treatment administration is that treatment time must be based on a patient's internal clock time (ICT), which can be uniquely offset relative to the 24-hour day-night cycle time (Zeitgeber time, or ZT; 
\citeauthor{Duffy2011}, \citeyear{Duffy2011}; \citeauthor{Huang2024}, \citeyear{Huang2024}; \citeauthor{Levi2007}, \citeyear{Levi2007}; \citeauthor{Ruben2019}, \citeyear{Ruben2019}). The offset of a patient's ICT relative to ZT, which is typically defined as the ZT at which a patient's melatonin levels start to rise under dim-light conditions (dim-light melatonin onset, or DLMO, time), can be determined from laboratory tests \citep{Lewy1999, Ruiz2020, Wittenbrink2018, Wright2013}. These laboratory tests require the collection and processing of multiple tissue samples from a patient over an extended period of time to monitor melatonin levels \citep{Kantermann2015, Reid2019, Kennaway2019}, which increases the cost of performing a study. The use of ZT rather than ICT for study analysis to reduce costs, however, would reduce the accuracy of study conclusions \citep{Gorczycaa2024, Gorczycad2024}.

To address the cost of DLMO time determination, research has increasingly focused on developing models that predict the ICT at which a distinct set of study-relevant biomarkers were measured \citep{Braun2018, Huang2024, Laing2017, Hughey2017}. While consistent improvements have been made in predicting the ICT of biomarker measurement over the past decade, these research efforts have largely relied on linear modelling frameworks for prediction \citep{Braun2018, Huang2024, Laing2017}. The use of linear models indicates that the potentially complex interactions among different biomarkers may not be adequately captured, and prediction accuracy could be further improved by accounting for these interactions. Further, these modelling frameworks have been assessed on data sets where biomarkers were measured longitudinally \citep{Archer2014, Braun2018, MllerLevet2013}, but these previous frameworks have focused on predicting ICT given biomarkers measured at a single time point.

In response to these two limitations, we present a recurrent neural network (RNN) modelling framework, which we refer to as LassoRNet (``Lasso Recurrent Network''), that is capable of predicting both ICT and DLMO time. A novel feature of LassoRNet is an extension of the Lasso, a data-driven technique that automates variable selection for a linear regression model \citep{Tibshirani1996}, to recurrent neural networks. The remainder of this paper is organized as follows. Section \ref{sec:2} gives an overview of the data, current state-of-the-art for ICT prediction, LassoRNet, and the experimental protocol for comparing current state-of-the-art to LassoRNet. Section \ref{sec:3} provides the results from this comparison. Finally, Section \ref{sec:4} discusses the significance of these experimental results.

\section{Materials and Methods} \label{sec:2}

In this section, it is assumed that tissue samples have been collected longitudinally from a cohort of $M$ people. The $i$-th person in this cohort provides $N_i$ tissue samples, which are processed to obtain the expression levels of $G$ different genes ($G$ different biomarkers). We define $\boldsymbol{X}_{i,j} = [X_{i,j,1} \ \ldots \ X_{i,j,G}]$ to be a $1\times G$ dimensional vector of normalized gene expression levels, where $X_{i,j,k}$ denotes the $j$-th measurement of the $k$-th gene's expression levels from the $i$-th person. In particular, these gene expression levels are normalized under a Z-score transform, where
\begin{align*}
    X_{i,j,k}=\frac{\tilde{X}_{i,j,k}- m_k}{s_k}. \label{eq:normalize}
\end{align*}
Here, $\tilde{X}_{i,j,k}$ represents the expression level of the $k$-th gene in the $j$-th sample from the $i$-th person recorded in a data set, $m_k$ denotes the mean expression level for the $k$-th gene across every person and sample, and $s_k$ denotes the corresponding standard deviation. We also define $\boldsymbol{Y}_{i}^{\dagger} = [Y^{\dagger}_{i,1} \ \ldots \ Y^{\dagger}_{i,N_i}]^T$ to be a $N_i \times 1$ vector of ZTs for the $i$-th person, with $Y^{\dagger}_{i,j}$ denoting the ZT at which $\boldsymbol{X}_{i,j}$ was collected. Finally, we define $\boldsymbol{Z} = [Z_{1} \ \ldots \ Z_{M}]^T$ to be an $M \times 1$ vector of DLMO times, where $Z_i$ denotes the $i$-th person's DLMO time, with $\boldsymbol{Y}_i^*=\boldsymbol{Y}^{\dagger}_i + Z_i$ denoting the corresponding ICTs of sample collection for the $i$-th person.

\subsection{Data Overview} \label{sec:2.1}

This paper considers three longitudinal transcriptome study data sets commonly used to benchmark the accuracy of ICT prediction methods \citep{Braun2018, Huang2024, Laing2017, Hughey2017}. In each of the corresponding studies that curated these data sets, two sets of tissue samples were obtained from each study participant over time: one set was used to derive the expression levels of multiple genes, and another set was used to determine a study participant's DLMO time. Each data set has already been processed and made publicly available, with each data set consisting of expression data from 7615 different genes \citep{Huang2024}.

The first data set will be referred to as the ``Archer data set.'' This data set was created during a study where 22 people participated in a sleep desynchrony protocol. In this protocol, each person stayed awake for 20 hours and slept for eight hours in a cycle over a total duration of 96 hours. Before and after the protocol, blood samples were collected from each person participating in the study once every four hours over a 24-hour period. The blood samples that were processed for gene expression levels yielded microarray data \citep{Archer2014}. We note that, in line with assumptions from previous benchmark studies, the blood samples collected before the protocol are treated as a separate set of samples from those collected after the protocol \citep{Braun2018, Laing2017, Huang2024}. In other words, we consider the pre-protocol and post-protocol samples as if they belong to two different people.

The second data set will be referred to as the ``Braun data set.'' This data set was developed to provide an additional benchmark for evaluating the performance of different ICT prediction methods. Specifically, this data set was created from blood samples gathered from 11 people with similar sleep schedules, health statuses, and ages. Each person provided a blood sample once every two hours over a 29-hour period. The blood samples that were processed for gene expression levels yielded next generation sequencing data \citep{Braun2018}.

The third data set will be referred to as the ``M\"{o}ller-Levet data set.'' This data set was curated during a study on the effects of how insufficient sleep affects gene expression. The study involved 24 people being placed under two interventions. The first intervention allowed each person to sleep for up to ten hours each night. The second intervention allowed each person to sleep for up to six hours each night. After each intervention, each person contributed a blood tissue sample once every three hours over a 30-hour period. The blood tissue samples that were processed for gene expression levels yielded microarray data \citep{MllerLevet2013}. Similar to the Archer data set, each set of intervention samples obtained from a single person is treated as if it were obtained from two different people.

While each of these data sets have been processed, we perform two additional processing steps that follow the protocol of other studies \citep{Gorczycaa2024, Gorczycab2024, Gorczycac2024}. First, data from people whose DLMO time could not be determined are excluded. Second, any genes with missing expression level measurements in a given sample population's data set are removed. A summary of how these processing steps affect each sample population's data set is provided in Table \ref{tab:data_summary}.

\subsection{Overview of ICT Prediction Methods} \label{sec:2.2}

In this study, we will consider three state-of-the-art methods for predicting the internal circadian time (ICT) at which a tissue sample is collected from a person. Each method aims to estimate a function $f(X_{i,j,1}, \dots, X_{i,j,G})$ that predicts ICT given expression levels of $G$ different genes. Specifically, the output of this function is a $1\times 2$ dimensional vector
\begin{align*} 
\begin{bmatrix} 
\hat{T}^*_{i,j,1} & 
\hat{T}^*_{i,j,2} 
\end{bmatrix} = f(X_{i,j,1}, \dots, X_{i,j,G}),
\end{align*} 
where the corresponding true values of $T^*_{i,j,1}$ and $T^*_{i,j,2}$ are defined as 
\begin{align*} 
\begin{bmatrix} 
T^*_{i,j,1} & 
T^*_{i,j,2} 
\end{bmatrix} = \begin{bmatrix} 
\sin\left( \frac{\pi Y^*_{i,j}}{12} \right) & 
\cos\left( \frac{\pi Y^*_{i,j}}{12} \right) 
\end{bmatrix}.
\end{align*} 
The reason these methods predict the vector $[T^*_{i,j,1} \ T^*_{i,j,2}]$ instead of directly predicting the ICT $Y^*_{i,j}$ is that time exhibits periodic behavior, or repeats on a 24-hour cycle. To illustrate why this transformation is useful, suppose $Y^*_{i,j} = 23$, which is near the end of a 24-hour cycle. If the model predicts $\hat{Y}^*_{i,j} = 1$, this prediction is only 2 hours away from the true value and this prediction is at the beginning of a 24-hour cycle. However, due to differences in location relative to the cycle boundary, the arithmetic difference between these two quantities is $Y^*_{i,j}-\hat{Y}^*_{i,j}=23 - 1 = 22$ hours, which is larger than the true difference of $\mathrm{2}$ hours. The transform of $Y^*_{i,j}$ to $[T^*_{i,j,1} \ T^*_{i,j,2}]$ is a mapping of time to account for predictions and true values being at different ends of a 24-hour interval when estimating this function.

It is emphasized that once the function $f(X_{i,j,1}, \dots, X_{i,j,G})$ is estimated, the ICT prediction $\hat{Y}^*_{i,j}$ can be obtained from the transform
\begin{align*}
\hat{Y}_{i,j}^*= \frac{12}{\pi} \left\{ \mathrm{atan2}\left(\hat{T}^*_{i,j,1}, \hat{T}^*_{i,j,2}\right) \mod 2\pi \right\}.
\end{align*}
Here, $\mathrm{atan2}(a, b)$ is the two-argument arctangent function with arguments $a$ and $b$. It is also noted that for the ICT prediction methods, we will also give consideration to an augmented version of this estimation task, where we instead aim to estimate a function
\begin{align*} 
\begin{bmatrix} 
\hat{T}^*_{i,j,1} & 
\hat{T}^*_{i,j,2} 
\end{bmatrix} = g(X_{i,j,1}, \dots, X_{i,j,G}, T^{\dagger}_{i,j,1}, T^{\dagger}_{i,j,1}),
\end{align*} 
with
\begin{align*} 
\begin{bmatrix} 
T^{\dagger}_{i,j,1} & 
T^{\dagger}_{i,j,2} 
\end{bmatrix} = \begin{bmatrix} 
\sin\left( \frac{\pi Y^{\dagger}_{i,j}}{12} \right) & 
\cos\left( \frac{\pi Y^{\dagger}_{i,j}}{12} \right) 
\end{bmatrix}.
\end{align*} 
The inclusion of $T^{\dagger}_{i,j,1}$ and $T^{\dagger}_{i,j,2}$ as additional input variables is due to the World Health Organization advocating that ZT be recorded during blood sample collection \citep{WHO2010}.

\subsubsection{Partial Least Squares Regression} \label{sec:2.2.1}

A challenge in predicting ICT from the expression levels of $G$ different genes is that $G$ can be larger than the total number of samples collected. Partial least squares regression (PLSR) addresses this issue by transforming both the input variables and the ICTs at which a tissue sample is collected into smaller dimensional latent spaces, where this transformation is algorithmically designed to maximize the covariance between the latent variables representing the input variables and ICT. Least squares regression is then performed on these output latent variables (\citealt[Section 3.5.2]{Hastie2009}; \citealt{Geladi1986}). 

When predicting ICT with a PLSR model, there are two hyper-parameters, or user-specified quantities, that need to be defined before regression. The first hyper-parameter is the dimension of the latent space for the input variables. The second hyper-parameter is a threshold parameter $K$, which takes the top $K$ input variables that have the largest weights associated with them when projected into the latent space. This paper considers the same hyper-parameter optimization protocol as the PLSR method used for ICT prediction, which considered projecting input variables to a 5 to 40 latent dimensional space (in increments of 5), and considering values of $K$ between 100 and 5000 \citep{Laing2017}.

\subsubsection{TimeMachine and TimeSignature} \label{sec:2.2.2}

The remaining two ICT prediction methods use penalized least squares regression to obrain a linear model for predicting ICT given gene expression levels. Specifically, when given gene expression levels as input variables, TimeSignature would assume that $[T^*_{i,j,1} \
T^*_{i,j,2}]$ can be modelled as
\begin{equation}
\begin{bmatrix} 
T^*_{i,j,1} &
T^*_{i,j,2}
\end{bmatrix} = \boldsymbol{\beta_0} + \boldsymbol{X}_{i,j}\boldsymbol{\beta} + \eta. \label{eq:lin_mod}
\end{equation}
Here, $\boldsymbol{\beta_0}$ denotes a $1\times 2$ dimensional intercept weight vector, $\boldsymbol{\beta}$ a $G\times 2$ dimensional weight matrix, and $\eta$ a $1\times 2$ dimensional vector of random noise. 

The parameters of the model in (\ref{eq:lin_mod}) are estimated by solving the optimization problem
\begin{equation}
\begin{aligned}
\boldsymbol{\hat{\beta}_0}, \boldsymbol{\hat{\beta}} &=\argmin_{(\boldsymbol{\beta_0}, \boldsymbol{\beta}) \in \mathbb{R}^{(G+1) \times 2}} \left\{ \frac{1}{2\left(\sum_{l=1}^MN_l\right)} \sum_{i=1}^{M}\sum_{j=1}^{N_i} \left\| \begin{bmatrix}T^*_{i,j,1} & T^*_{i,j,2} \end{bmatrix} - \boldsymbol{\beta_0} - \boldsymbol{X}_{i,j}\boldsymbol{\beta} \right\|_2^2\right\} \\
& \quad \quad \quad \quad \quad \quad \quad + \lambda \left[ \left. \left\{\frac{(1-\alpha)}{2}\sum_{k=1}^{G} (\boldsymbol{\beta}_{k,1}^2 + \boldsymbol{\beta}_{k,2}^2)^2\right\}  + \left(\alpha \sum_{k=1}^{G} \sqrt{\boldsymbol{\beta}_{k,1}^2 + \boldsymbol{\beta}_{k,2}^2} \right) \right. \right]. 
\end{aligned}\label{eq:elastic_net} 
\end{equation}
In the objective function of (\ref{eq:elastic_net}), the first term corresponds to the optimization problem used in least squares regression. The second term penalizes the magnitude of the parameter estimates. To clarify, there are two hyper-parameters in this penalty term, $\lambda$ and $\alpha$. The term $\lambda$ represents the severity of the penalty applied (a larger value of $\lambda$ decreases the magnitude of each element in $\boldsymbol{\beta}$ towards zero). The term $\alpha$ represents how this penalty is distributed between the quantities $\sum_{k=1}^{G} (\boldsymbol{\beta}_{k,1}^2 + \boldsymbol{\beta}_{k,2}^2)^2$ and $\sum_{k=1}^G\sqrt{\boldsymbol{\beta}_{k,1}^2+\boldsymbol{\beta}_{k,2}^2}$. We follow the same hyper-parameter optimization protocol specified in the methodology for TimeMachine and TimeSignature for identifying $\lambda$ and $\alpha$.

The key difference between TimeMachine and TimeSignature is the number of genes considered. TimeMachine first selects 37 genes before solving the optimization problem in (\ref{eq:elastic_net}), whereas TimeSignature uses every gene available \citep{Braun2018, Huang2024}. 

\subsection{LassoRNet Overview} \label{sec:2.3}

\subsubsection{Background on Bidirectional Recurrent Neural Networks} \label{sec:2.3.1}

Recurrent neural networks (RNN) describe a modelling framework for representing longitudinal data. Specifically, a RNN computes a sequence of hidden vectors for the $i$-th person, denoted as $\boldsymbol{h}_i = \{h_{i,1}, \ \ldots, \ h_{i,N_i}\}$, as well as a sequence of output vectors, denoted as $\boldsymbol{o}_i = \{o_{i,1}, \ \ldots, \ o_{i,N_i}\}$. For data obtained from the $i$-th person, these two sequences are derived by iterating over the following equations from $j=1$ to $j=N_i$:
\begin{align*}
    h_{i,j} &= \mathcal{H}(\boldsymbol{X}_{i,j}W_{X,h} + h_{i,j-1}W_{h,h}+b_h), \\
    o_{i,j} &= h_{i,j}W_{h,o}+b_o.
\end{align*}
Here, $W_{X,h}$, $W_{h,h}$ and $W_{h,o}$ represent weight matrices \citep{Graves2013}. Specifically, $W_{X,h}$ represents the weight matrix that interacts with gene expression data from the $j$-th sample of the $i$-th person, or $\boldsymbol{X}_{i,j}$, to produce a hidden vector $h_{i,j}$; $W_{h,h}$ the weight matrix that interacts with the hidden vector $h_{i,j}$; and $W_{h,o}$ the weight matrix utilized to compute the output vector $o_{i,j}$, which would be modelled further for predicting ICT. In addition to these three weight matrices, $b_h$ and $b_o$ represent intercept weight vectors that are used in computing $h_{i,j}$ and $o_{i,j}$, respectively; and $\mathcal{H}(U)$ represents a hidden layer function with respect to an argument $U$.

While $\mathcal{H}(U)$ can be defined as any function, this paper defines $\mathcal{H}(U)$ to be a long short-term memory (LSTM) unit, which is popular for RNNs \citep{Graves2012, Hochreiter1997, Gers2003}. A LSTM unit is defined by the composite functions
\begin{equation}
\begin{aligned}
e_{i,j} &= \sigma(\boldsymbol{X}_{i,j}W^{(1)} + b^{(1)} + h_{i,j-1}W^{(5)}  + b^{(5)}), \\
p_{i,j} &= \sigma(\boldsymbol{X}_{i,j}W^{(2)} + b^{(2)} + h_{i,j-1}W^{(6)}  + b^{(6)}), \\
a_{i,j} &= \tanh(\boldsymbol{X}_{i,j}W^{(3)} + b^{(3)} + h_{i,j-1}W^{(7)} + b^{(7)}), \\
o_{i,j} &= \sigma(\boldsymbol{X}_{i,j}W^{(4)} + b^{(4)} + h_{i,j-1}W^{(8)} + b^{(8)}), \\
c_{i,j} &= p_{i,j} \odot c_{i,j-1} + e_{i,j} \odot a_{i,j}, \\
h_{i,j} &= o_{i,j} \odot \tanh(c_{i,j}).
\end{aligned} \label{eq:lstm}
\end{equation}
Here, $\sigma(U) = [1/\{1+\exp(-U_1)\} \ \ldots \ 1/\{1+\exp(-U_K)\}]$ represents an element-wise logistic sigmoid function with respect to a $K$-dimensional vector $U$, and $\tanh(U)$ an element-wise hyperbolic tangent function. Further, $\odot$ represents the Hadamard product \citep{pytorch2021}.

A limitation of conventional RNNs is their reliance solely on past context. When every tissue sample is processed prior to prediction, it is feasible to use both past and future context provided by each sample relative to the other samples. Bidirectional RNNs (BRNNs) address this by processing data in both directions using two distinct hidden layers, which then feed into the same output layer \citep{Schuster1997}. Specifically, a BRNN computes a forward hidden sequence $\boldsymbol{\overrightarrow{h}_i} = \{ \overrightarrow{h}_{i,1},\ldots, \overrightarrow{h}_{i,N_i}\}$, a backward hidden sequence $\boldsymbol{\overleftarrow{h}_i} = \{ \overleftarrow{h}_{i,1},\ldots, \overleftarrow{h}_{i,N_i}\}$, and an output sequence $\boldsymbol{o_i} = \{o_{i,1}, \ \ldots, \ o_{i,N_i}\}$ for the $i$-th person by iterating through the equations
\begin{align*}
    \overrightarrow{h}_{i,j} &= \mathcal{H}(\boldsymbol{X}_{i,j}W_{X,\overrightarrow{h}} + \overrightarrow{h}_{i,j-1}W_{\overrightarrow{h},\overrightarrow{h}}+b_{\overrightarrow{h}}), \\
    \overleftarrow{h}_{i,j} &= \mathcal{H}(\boldsymbol{X}_{i,j}W_{X,\overleftarrow{h}} + \overleftarrow{h}_{i,j-1}W_{\overleftarrow{h},\overleftarrow{h}}+b_{\overleftarrow{h}}), \\
    o_{i,j} &= \overrightarrow{h}_{i,j}W_{\overrightarrow{h},o} + \overleftarrow{h}_{i,j}W_{\overleftarrow{h},o}+b_o.
\end{align*}
Combining BRNNs with a LSTM unit results in bidirectional LSTMs, which can leverage long-range context in both input directions \citep{Graves2005}.

\subsubsection{Variable Selection for LassoRNet} \label{sec:2.3.2}

We propose a new approach that extends the optimization problem used to estimate the weights for TimeMachine and TimeSignature in (\ref{eq:elastic_net}) to estimating the weights of bidirectional LSTMs. This new approach defines a neural network architecture for predicting $[T^*_{i,j,1}  \ T^*_{i,j,2} ]$ as
\begin{align*} 
\begin{bmatrix} 
\hat{T}^*_{i,j,1} & 
\hat{T}^*_{i,j,2} 
\end{bmatrix} &= o_{i,j}\theta + [\boldsymbol{X}_{i,j} \ T^{\dagger}_{i,j,1} \ T^{\dagger}_{i,j,2}]\boldsymbol{\beta}.
\end{align*} 
Here, $o_{i,j}$ represents the $j$-th output vector for the $i$-th person from (\ref{eq:lstm}) for a nonlinear bidirectional LSTM when given both gene expression levels $\boldsymbol{X}_{i,j}$ and transformed ZT, or $[T^{\dagger}_{i,j,1} \ T^{\dagger}_{i,j,2}]$, as input variables. Further, $[\boldsymbol{X}_{i,j} \ T^{\dagger}_{i,j,1} \ T^{\dagger}_{i,j,2}]\boldsymbol{\beta}$, which is referred to as a residual connection in machine learning literature
\citep{He2016, Lemhadri2021}, can be interpreted as the linear model specified by TimeMachine and TimeSignature.

The weights of the linear component $[\boldsymbol{X}_{i,j} \ T^{\dagger}_{i,j,1} \ T^{\dagger}_{i,j,2}]\boldsymbol{\beta}$ and the weights of the nonlinear component $o_{i,j}\theta$ are estimated jointly under the optimization problem 
\begin{equation}
\begin{aligned} 
\min_{\beta, \theta, \boldsymbol{W}} & \ \left\{\frac{1}{2\left(\sum_{l=1}^MN_l\right)} \sum_{i=1}^M\sum_{j=1}^{N_i}||\begin{bmatrix}T_{i,j,1}^* & T_{i,j,2}^*\end{bmatrix} -o_{i,j}\theta + [\boldsymbol{X}_{i,j} \ T^{\dagger}_{i,j,1} \ T^{\dagger}_{i,j,2}]\boldsymbol{\beta}||_2^2\right\} \\
&\quad \quad + \lambda \sum_{k=1}^{G+2} \sqrt{\boldsymbol{\beta}_{k,1}^2 + \boldsymbol{\beta}_{k,2}^2}\\ 
\mathrm{subject\  to} &\   
|\{\overleftarrow{W}^{(m)}\}_{k,l}| \leq \tau \sqrt{\boldsymbol{\beta}_{k,1}^2+\boldsymbol{\beta}_{k,2}^2}, \quad \forall k, \ \forall l, \mathrm{and} \  m \in \{1,2,3,4\}, \\
&\   
|\{\overrightarrow{W}^{(m)}\}_{k,l}| \leq \tau \sqrt{\boldsymbol{\beta}_{k,1}^2+\boldsymbol{\beta}_{k,2}^2}, \quad \forall k, \ \forall l, \ \mathrm{and} \  m \in \{1,2,3,4\},
\end{aligned} \label{eq:lassornet}
\end{equation}
Here, $\boldsymbol{W}$ represents the weights of the bidirectional LSTM, while the weights $\{\overleftarrow{W}_1,\ldots, \overleftarrow{W}_4,\overrightarrow{W}_1,\ldots, \overrightarrow{W}_4\}$ are those from the bidirectional LSTM that interact with the input variables, $[\boldsymbol{X}_{i,j}, T^{\dagger}_{i,j,1}, T^{\dagger}_{i,j,2}]$, to produce $o_{i,j}$. The motivation for (\ref{eq:lassornet}) is to select relevant input variables (such as the $G$ genes) for the entire neural network. This input variable selection is due to the constraints 
\begin{align*} 
\   
|\{\overleftarrow{W}^{(m)}\}_{k,l}| &\leq \tau \sqrt{\boldsymbol{\beta}_{k,1}^2+\boldsymbol{\beta}_{k,2}^2}, \quad \forall k, \ \forall l, \ \mathrm{and} \  m \in \{1,2,3,4\}, \\
\   
|\{\overrightarrow{W}^{(m)}\}_{k,l}| &\leq \tau \sqrt{\boldsymbol{\beta}_{k,1}^2+\boldsymbol{\beta}_{k,2}^2}, \quad \forall k, \ \forall l, \ \mathrm{and} \  m \in \{1,2,3,4\},
\end{align*} 
which enforce all weights that interact with the $k$-th input variable in a bidirectional LSTM unit to be no larger than the corresponding linear weights $[\boldsymbol{\beta}_{k,1} \ \boldsymbol{\beta}_{k,2}]$ used to predict ICT. To clarify, this optimization problem is designed such that the nonlinear contribution for the $k$-th input variable is based on the linear contribution of $[\boldsymbol{\beta}_{k,1} \ \boldsymbol{\beta}_{k,2}]$, with the magnitude of the elements of $\beta$ controlling the sparsity of the entire neural network. When $\sqrt{\boldsymbol{\beta}_{k,1}^2+\boldsymbol{\beta}_{k,2}^2} = 0$, the $k$-th input variable does not provide any contribution to ICT prediction. Further, when the quantity $\tau$ in the constraints are set equal to zero, the optimization problem (\ref{eq:lassornet}) is the same as that used for augmented versions of TimeMachine and TimeSignature from (\ref{eq:elastic_net}) where $\alpha=1$, which is known as the Lasso for data-driven variable selection \citep{Tibshirani1996}. It is noted that as $\tau$ increases, the influence of $\boldsymbol{\beta}$ for sparsity in the number of selected variables decreases, and an unconstrained bidirectional LSTM is obtained as $\tau$ converges to infinity. 

The optimization problem in (\ref{eq:lassornet}) requires a researcher to specify two hyper-parameters: the penalty parameter $\lambda$, which controls sparsity in the number of selected input variables, and the hierarchy parameter $\tau$, which adjusts the contribution of the linear component from the residual connection relative to the nonlinear component from the bidirectional LSTM. In Appendix \ref{app:A}, we describe our procedure for solving this optimization problem given specified values for these hyper-parameters.

\subsection{Extending Each Method to Predict DLMO Time}

While each of the methods have been designed to predict ICT given biomarker measurements, the underlying offset of a person's ICT relative to ZT is dictated by that person's DLMO time. To enable identification of DLMO time, we propose an extension for each of the ICT methods previously described. Specifically, for PLSR, TimeMachine, and TimeSignature, we define DLMO time prediction as
\begin{align}
    \hat{Z}_i &= (\hat{\mathrm{Y}}^*_{i,\tilde{j}_i} - Y^{\dagger}_{i,\tilde{j}_i}) \mod 24. \label{eq:old_find}
\end{align}
Here, $\tilde{j}_i$ is the index of the closest recorded ZT for the $i$-th person relative to the ZT at which ICT prediction error was minimized across every person participating in a study. For LassoRNet, we instead estimate the weights
\begin{align}
    \alpha^* &=\argmin_{\alpha} \ \sum_{i=1}^M\left(Z_i - \begin{bmatrix}(\hat{\mathrm{Y}}^*_{i,\tilde{j}_i}-\hat{\mathrm{Y}}^{\dagger}_{i,\tilde{j}_i}) & (\hat{\mathrm{Y}}^*_{i,\tilde{j}_i+1}-\hat{\mathrm{Y}}^{\dagger}_{i,\tilde{j}_i+1}) & (\hat{\mathrm{Y}}^*_{i,\tilde{j}_i+2}-\hat{\mathrm{Y}}^{\dagger}_{i,\tilde{j}_i+2}) \end{bmatrix}\alpha\right)^2 \label{eq:new_find}
\end{align}
and define the $i$-th person's DLMO time prediction as
\begin{align*}
    \hat{Z}_i = \begin{bmatrix}(\hat{\mathrm{Y}}^*_{i,\tilde{j}_i}-\hat{\mathrm{Y}}^{\dagger}_{i,\tilde{j}_i}) & (\hat{\mathrm{Y}}^*_{i,\tilde{j}_i+1}-\hat{\mathrm{Y}}^{\dagger}_{i,\tilde{j}_i+1}) & (\hat{\mathrm{Y}}^*_{i,\tilde{j}_i+2}-\hat{\mathrm{Y}}^{\dagger}_{i,\tilde{j}_i+2}) \end{bmatrix}\alpha^*
\end{align*}
To clarify, we instead identify a sequence of three time points and a corresponding weight vector $\alpha^*$ that minimize the mean squared error in predicting DLMO time, where the optimal ZT for minimizing this error is identified in the same manner as PLSR, TimeMachine, and TimeSignature. We consider three sequential time points due to its prior interest in evaluating ICT prediction frameworks \citep{Laing2017}.

\subsection{Model Training and Assessment} \label{sec:2.4}

The objective of this study is to evaluate different methods for their ability to predict ICT and DLMO time. To provide a brief summary of the training and assessment protocol given a data set, it is first divided into three parts: a training data set (consisting of samples from $0.4M$ people), a validation data set (consisting of samples from $0.3M$ people), and a test data set (consisting of samples from $0.3M$ people). The training data set is used to obtain a model given hyper-parameters specified before the training process. To optimize these hyper-parameters, the model obtained is then assessed on the validation data set. This paper selects the model that minimizes the mean squared error in ICT prediction on the validation set, which has been used as the selection criterion in prior studies \citep{Laing2017, Braun2018, Huang2024}. It is emphasized that PLSR, TimeMachine, and TimeSignature models were trained in the R statistical software \citep{R2021} using the same hyper-parameter search procedure from their corresponding studies. For LassoRNet, the hyper-parameters were optimized using a random search \citep{Bergstra2012}. In this study, 50 random searches are performed for LassoRNet using PyTorch \citep{pytorch2021}. 

For model assessment in ICT prediction, we give consideration to two quantities commonly reported for this task. The first quantity is the median absolute error ($\mathrm{MAE}_{\mathrm{ICT}}$), or
\begin{align*}
    \mathrm{MAE}_{\mathrm{ICT}} &= \median_{(i,j)\in \{\mathrm{Test \ Data \ Set}\}}\min\left\{|(Y^*_{i,j} - \hat{Y}^*_{i,j}) \mod 24|, 24 - |(Y^*_{i,j} - \hat{Y}^*_{i,j}) \mod 24|\right\}.
\end{align*}
The second quantity is defined as
\begin{align*}
    \mathrm{AUC}_{\mathrm{ICT}} &= 1-\frac{1}{12}\int_0^{12}\left[ \frac{1}{\left(\sum_{l=1}^MN_l\right)}\sum_{i=1}^M\sum_{j=1}^{N_i} \left\{\mathbbm{1}_{\delta(Y_{i,j}, \hat{Y}_{i,j})} (U)\right\}\right]dU,
\end{align*}
where
\begin{align*}
    \mathbbm{1}_{\delta(Y_{i,j}, \hat{Y}_{i,j})} (U)&=
\begin{cases}
  1, & \mathrm{if} \  \min\left\{|(Y^*_{i,j} - \hat{Y}^*_{i,j}) \mod 24|, 24 - |(Y^*_{i,j} - \hat{Y}^*_{i,j}) \mod 24|\right\} > U, \\
  0, & \mathrm{otherwise.}
\end{cases}
\end{align*}
This second performance measure is referred to as the ``area under the curve'' in ICT prediction tasks, and can be interpreted as a normalized form of the mean absolute error \citep{Braun2018, Huang2024}. Specifically, when $\mathrm{AUC}_{\mathrm{ICT}}=1$, then a model's predictions equal the true ICTs (there is no prediction error). Further, when $\mathrm{AUC}_{\mathrm{ICT}}=0.5$, there are two common interpretations of this performance measure. The first common interpretation is that the model's predictions are random and uniformly distributed over a 24-hour interval. The second common interpretation is that if the model's ICT predictions are a constant quantity, then the true ICT values are uniformly distributed over a 24-hour interval. It is noted that corresponding performance measures for DLMO time were obtained in the same manner, where determination of each $\tilde{j}_i$ in (\ref{eq:old_find}) and (\ref{eq:new_find}) as well as computation $\hat{\alpha}$ in (\ref{eq:new_find}) was performed with data from the validation data set.

\section{Results} \label{sec:3}

Table \ref{tab:res1} displays the results for ICT prediction. We find that LassoRNet outperforms partial least squares regression (PLSR), TimeMachine, and TimeSignature in predicting ICT on the out-of-sample test data set. In particular, LassoRNet maintains at least a 30 minute reduction of the median absolute error in ICT prediction when compared to these three methods (TimeSignature obtains the closest performance to LassoRNet in terms of the median absolute error when compared on the M\"{o}ller-Levet data set). Table \ref{tab:res1} also presents assessments of ``augmented'' versions of PLSR, TimeMachine, and TimeSignature, where each of these methods include sine and cosine transforms of Zeitgeber time (ZT) as additional predictors. While LassoRNet outperforms all three of these methods given augmented data, the augmented versions of these methods obtain consistently strong performance relative to their ``baseline'' (not augmented) counterparts. A noticeable example of this occurs in the Archer data set assessment, where the median absolute error for the augmented methods are at least one hour less than their corresponding baselines.

Table \ref{tab:res2} displays corresponding results for dim-light melatonin onset (DLMO) time prediction. We find that LassoRNet further improves in DLMO time prediction accuracy, obtaining median absolute errors between 30 and 40 minutes. The augmented versions of PLSR, TimeMachine, and TimeSignature also outperform their baseline counterparts. A notable example of this is TimeMachine for the Braun data set, where the augmented version obtains an approximately 2 hour reduction in the median absolute error equal to approximately 44 minutes.

\section{Discussion} \label{sec:4}

In this paper, we propose a recurrent neural network (RNN) framework for predicting a person's internal circadian time (ICT) as well as their offset between ICT relative to the 24-hour day-night cycle time (Zeitgeber time, or ZT; \citeauthor{Lewy1999}, \citeyear{Lewy1999}; \citeauthor{Ruiz2020}, \citeyear{Ruiz2020}; \citeauthor{Wittenbrink2018}, \citeyear{Wittenbrink2018}; \citeauthor{Wright2013}, \citeyear{Wittenbrink2018}). The development of this framework is motivated by the cost of determining the offset of a person's ICT relative to ZT. Specifically, the laboratory tests used for determining this offset requires a technician to collect multiple tissue samples over time and extrapolate the ZT at which melatonin levels begin to rise under dim-light conditions (dim-light melatonin onset, or DLMO, time; \citeauthor{Kantermann2015}, \citeyear{Kantermann2015}; \citeauthor{Reid2019}, \citeyear{Reid2019}; \citeauthor{Kennaway2019}, \citeyear{Kennaway2019}; \citeauthor{Kennaway2023}, \citeyear{Kennaway2023}). We find that the proposed RNN framework noticeably improves over state-of-the-art, obtaining a median absolute error between 30 to 40 minutes when predicting DLMO time.

While this framework improves over current state-of-the-art for ICT and DLMO time prediction, this improvement requires the collection of multiple tissue samples over time. In scenarios where only a single sample is collected from each person, this may not be feasible. For these scenarios, we recommend augmenting methods for ICT prediction by including information about the ZT of biomarker measurement, which we have shown consistently improves ICT prediction accuracy. This improvement is particularly evident when comparing augmented versions of TimeMachine and TimeSignature obtained from the Archer data set to their baseline counterparts (Table \ref{tab:res1}), as these baseline counterparts did not select any genes for prediction (an intercept model was output from computation). This result is not surprising given the additional context that the experimental conditions from \citeauthor{Archer2014}, \citeyear{Archer2014} led to a substantive changes in the behavior of each gene's expression levels over time. This recommendation also aligns with guidelines from \citeauthor{WHO2010}, \citeyear{WHO2010}, which advocates that the ZT of biomarker measurement is recorded for blood tissue samples.

\section*{Acknowledgements}
This work was performed under the auspices of the US Department of Energy by Lawrence Livermore National Laboratory under Contract DE-AC52-07NA27344 (LLNL-JRNL-2004512-DRAFT).

\section*{Conflict of Interest}
The authors declare no potential conflict of interests.

\section*{Data Availability Statement}

The authors will provide access to the software developed to reproduce the results presented in this paper upon reasonable request.

\clearpage
\newpage

\begin{table*}[!h]
    \caption{Summary of each data set before and after processing.}
    \label{tab:data_summary}
      \centering
		\resizebox{1.0\textwidth}{!}{
    \begin{tabular}{|c|c|c|c|c|}
    \hline
   & \textbf{Processing Stage} & \textbf{Archer} & \textbf{Braun} & \textbf{M\"{o}ller-Levet} \\
   \hline
   \multirow{2}{*}{\textbf{Number of Unique People}} & Before & 22 & 11 & 24  \\
   & After & 19 & 11 & 20 \\
   \hline 
   \multirow{2}{*}{\textbf{Total Sample Size}} & Before & 286 & 153 & 427 \\
   & After & 258 & 153 & 355 \\
   \hline 
   \multirow{2}{*}{\textbf{Number of Genes}} & Before & 7615 & 7615 & 7615 \\
   & After & 3689 & 7615 & 7615 \\
   \hline 
   \multirow{2}{*}{\textbf{Number of Genes (TimeMachine)}} & Before & 37 & 37 & 37 \\
   & After & 22 & 37 & 37 \\
         \hline 
    \end{tabular}}
\end{table*}

\clearpage
\newpage

\begin{table*}[!h]
	\caption{Model performance on out-of-sample test data for ICT prediction. Here, a ``$\ddagger$'' in the superscript of a method's performance measure indicates that the none of the input variables were selected for prediction.} \label{tab:res1}
 \centering
		\begin{tabular}{|c|c|c|c|c|c|c|c|c|c|c|c|}
			\hline
    \multicolumn{3}{|c|}{\textbf{\textit{Archer}}} \\
            \hline 
   Method  & Median Absolute Error & Area Under Curve \\ 
   \hline
   LassoRNet & $\boldsymbol{1.000}$ & $\boldsymbol{0.902}$ \\
   PLSR & $3.181$ & $0.685$ \\
   TimeMachine & $ \ \ 5.176^{\ddagger}$ & $ \ \ 0.545^{\ddagger}$ \\
   TimeSignature & $ \ \ 5.176^{\ddagger}$ & $ \ \ 0.545^{\ddagger}$ \\ 
   PLSR (Augmented) & $2.012$ & $0.787$ \\
   TimeMachine (Augmented) & $1.189$ & $0.885$ \\ 
   TimeSignature (Augmented) & $1.239$ &  $0.883$ \\
   \hline
    \multicolumn{3}{|c|}{\textbf{\textit{Braun}}} \\
    \hline 
   Method  & Median Absolute Error & Area Under Curve \\ 
   \hline
   LassoRNet & $\boldsymbol{1.058}$ & $\boldsymbol{0.877}$ \\
   PLSR & $2.113$ & $0.772$ \\
   TimeMachine & $1.795$ & $0.760$ \\ 
   TimeSignature & $4.154$ &  $0.581$ \\
   PLSR (Augmented) & $2.550$ & $0.741$ \\
   TimeMachine (Augmented) & $1.470$ & $0.875$ \\
   TimeSignature (Augmented) & $1.409$ & $\boldsymbol{0.877}$ \\
   \hline
   \multicolumn{3}{|c|}{\textbf{\textit{M\"{o}ller-Levet}}} \\
   \hline 
   Method  & Median Absolute Error & Area Under Curve \\ 
   \hline
   LassoRNet & $\boldsymbol{0.925}$ & $\boldsymbol{0.901}$ \\
   PLSR & $1.924$ & $0.819$ \\
   TimeMachine & $1.794$ & $0.806$ \\ 
   TimeSignature & $1.504$ &  $0.816$ \\
   PLSR (Augmented) & $2.747$ & $0.693$ \\
   TimeMachine (Augmented) & $1.801$ & $0.849$ \\ 
   TimeSignature (Augmented) & $1.765$ & $0.849$ \\
\hline
\end{tabular}
\end{table*}

\clearpage
\newpage

\begin{table*}[!h]
	\caption{Model performance on out-of-sample test data for DLMO time prediction. Here, a ``$\ddagger$'' in the superscript of a method's performance measure indicates that the none of the input variables were selected for prediction.} \label{tab:res2}
 \centering
		\begin{tabular}{|c|c|c|c|c|c|c|c|c|c|c|c|}
			\hline
    \multicolumn{3}{|c|}{\textbf{\textit{Archer}}} \\
            \hline 
   \textbf{Method}  & \textbf{Median Absolute Error} & \textbf{Area Under Curve} \\ 
   \hline
   LassoRNet & $\boldsymbol{0.514}$ & $\boldsymbol{0.912}$ \\
   PLSR & $3.253$ & $0.726$ \\
   TimeMachine & $ \ \ 1.093^{\ddagger}$ & $ \ \ 0.905^{\ddagger}$ \\
   TimeSignature & $ \ \ 1.093^{\ddagger}$ & $ \ \ 0.905^{\ddagger}$ \\ 
   PLSR (Augmented) & $2.493$ & $0.748$ \\
   TimeMachine (Augmented) & $1.325$ & $0.887$ \\ 
   TimeSignature (Augmented) & $1.340$ & $0.889$  \\
   \hline
    \multicolumn{3}{|c|}{\textbf{\textit{Braun}}} \\
    \hline 
   \textbf{Method}  & \textbf{Median Absolute Error} & \textbf{Area Under Curve} \\ 
   \hline
   LassoRNet & $\boldsymbol{0.500}$ & $\boldsymbol{0.971}$ \\
   PLSR & $1.030$ & $0.869$ \\
   TimeMachine & $2.528$ & $0.713$ \\ 
   TimeSignature &  $3.618$ & $0.701$ \\
   PLSR (Augmented) & $1.242$ & $0.882$ \\
   TimeMachine (Augmented) & $0.730$ & $0.932$ \\
   TimeSignature (Augmented) & $1.770$ & $0.840$  \\
   \hline
   \multicolumn{3}{|c|}{\textbf{\textit{M\"{o}ller-Levet}}} \\
   \hline 
   \textbf{Method}  & \textbf{Median Absolute Error} & \textbf{Area Under Curve} \\ 
   \hline
   LassoRNet & $\boldsymbol{0.673}$ & $\boldsymbol{0.915}$ \\
   PLSR & $2.366$ & $0.797$ \\
   TimeMachine & $3.194$ & $0.724$ \\ 
   TimeSignature & $0.907$ & $0.887$ \\
   PLSR (Augmented) & $3.108$ & $0.686$ \\
   TimeMachine (Augmented) & $1.316$ & $0.860$ \\ 
   TimeSignature (Augmented) & $1.319$ & $0.860$ \\
\hline
\end{tabular}
\end{table*}

\clearpage
\newpage

\appendix

\section{Overview of LassoRNet Training} \label{app:A}
We utilize proximal gradient descent to solve the optimization problem for LassoRNet from (\ref{eq:lassornet}). To briefly provide some background, proximal gradient descent is an extension of gradient descent, which is a common paradigm for training neural networks \citep[Section 4.3]{Goodfellow2016}. The optimization problem that motivates proximal gradient descent is defined as
\begin{align}
\min_{\gamma} \ f(\gamma) + g(\gamma). \label{eq:prox_basic}
\end{align}
Here, $f(\gamma)$ is a differentiable and smooth function with respect to the weights $\gamma$, while $g(\gamma)$ is a potentially non-differentiable, but convex function with respect to $\gamma$. The key idea behind proximal gradient descent is to decompose minimizing the objective function $f(\gamma) + g(\gamma)$ into two steps that are repeated until convergence: first, a gradient descent step for the smooth function $f(\gamma)$, followed by a ``proximal'' step for the function $g(\gamma)$.

Given the objective function $f(\gamma)+g(\gamma)$ as an example, we would perform the gradient descent step by computing the gradient of the smooth function $f(\gamma)$ with respect to $\gamma$, or $\nabla f(\gamma)$, and updating the current solution to this optimization problem after the $k$-th proximal step, which we denote as $\gamma^{(\mathrm{old})}$, using the equation
\begin{align*}
   \gamma^{(\mathrm{new})} = \gamma^{(\mathrm{old})} - \alpha \nabla f(\gamma^{(\mathrm{old})}).
\end{align*}
Here, $\alpha$ is a user-specified ``step size'', and $\gamma^{(\mathrm{new})}$ is the quantity obtained after the $k+1$-th (current) gradient descent step. The $k+1$-th proximal step then obtains a quantity $\rho^{(\mathrm{new})}$ by solving the optimization problem
\begin{align*}
   \rho^{(\mathrm{new})} = \argmin_{\rho} \ g(\rho) + \frac{1}{2} ||\gamma^{(\mathrm{new})} - \rho||^2_2.
\end{align*}
The quantity $\rho^{(\mathrm{new})}$ is a correction of $\gamma^{(\mathrm{new})}$ and solves the optimization problem in (\ref{eq:prox_basic}) after convergence \citep[Chapter 10]{Beck2017}. For each new update step performed, $\gamma^{(\mathrm{old})}$ is set equal to $\rho^{(\mathrm{new})}$, and proximal gradient descent outputs $\rho^{(\mathrm{new})}$ at convergence.

The derivation of the proximal step for LassoRNet extends the derivation presented by \citeauthor{Lemhadri2021}, \citeyear{Lemhadri2021}. Specifically, we consider solving a generalized form of the optimization problem presented in (\ref{eq:lassornet}),
\begin{align*}
    \min_{b, W_{1},\ldots,W_{L}} \quad & \frac{1}{2}||v-b||_2^2+\lambda||b||_2+\sum_{l=1}^L\left(||U_{l}-W_{l}||^2_2 +\bar{\lambda}||W_l||_1\right) \label{eq:optim}\\
     \ \ \ \ \ \mathrm{subject \ to}   \quad &  ||W_l||_{\infty}\leq \tau||b||_2 \ \ \forall l\in [L]. \nonumber
\end{align*}
In the context of the optimization problem from (\ref{eq:lassornet}), $v$ would represent the weights $[\boldsymbol{\beta}_{k,1} \ \boldsymbol{\beta}_{k,2}]$ after a gradient descent update. Further, if $L=8$, we would have
\begin{align*}
    \{U_1,\ldots, U_8\} = \{\overleftarrow{W}^{(1)}_{k,\cdot}, \ldots, \overleftarrow{W}^{(4)}_{k,\cdot}, \overrightarrow{W}^{(1)}_{k,\cdot}, \ldots, \overrightarrow{W}^{(4)}_{k,\cdot}\}
\end{align*}
be the set of corresponding weight vectors that interact with the $k$-th input variable from a bidirectional LSTM after a gradient descent step. 

Similar to \citeauthor{Lemhadri2021}, \citeyear{Lemhadri2021}, the closed-form expressions derived for this proximal step in Appendix \ref{as:app1.1} rely on a vector $\boldsymbol{s} = [s_1 \ \ldots \ s_L]$, which does not admit a closed-form expression. To reconcile this, we perform a global search procedure for identifying the optimal vector $\boldsymbol{s}$, which is summarized in Appendix \ref{as:app1.2}.

\subsection{Derivation of the Proximal Step for LassoRNet} \label{as:app1.1}

\begin{definition} \label{def1}
    The soft-thresholding operator $\mathcal{S}_{\lambda}(V) = \mathrm{sign}(V)\mathrm{max}(|Z|-\lambda, 0)$ given a univariate argument $V$.
\end{definition}

\begin{definition} \label{def2}
Suppose $V\in \mathbb{R}^M$. The $i$-th sorted coordinate of $V$ is denoted as $V_{(i)}$, and the sorted coordinate system satisfies $|V_{(1)}|\geq |V_{(2)}|\geq...\geq |V_{(M)}|$. 
\end{definition}

\begin{definition}
The subset of natural numbers $\{1,\ldots,V\}$ is denoted by $[V]$.
\end{definition}

\begin{theorem} \label{thm1}
Let $v \in \mathbb{R}^k$ and $U_{l}\in \mathbb{R}^K$, $l=1,\ldots,L$, denote fixed vectors. Consider the optimization problem
\begin{equation}
\begin{aligned}
    b^*, W^*_{1},\ldots,W^*_{L} = \argmin_{b, W_{1},\ldots,W_{L}} \quad & \frac{1}{2}||v-b||_2^2+\lambda||b||_2+\sum_{l=1}^L\left(||U_{l}-W_{l}||^2_2 +\bar{\lambda}||W_l||_1\right) \\
     \ \ \ \ \ \mathrm{subject \ to}   \quad &  ||W_l||_{\infty}\leq \tau||b||_2 \ \ \forall l\in [L]. 
\end{aligned} \label{eq:optim}
\end{equation}
The minimizers of this optimization problem are given by
\begin{align*}
    W^{*}_l &= \mathrm{sign}(U_l)\mathrm{min}\left\{\tau||b^*||_2, \mathcal{S}_{\bar{\lambda}}(U_l)\right\}, \\
    b^* &= \left(\frac{1}{1+\tau^2\sum_{l \in [L]}s_l}\right)\mathrm{max}\left(1-\frac{a_{\boldsymbol{s}}}{||v||_2}, 0\right)v,
\end{align*}
where $\boldsymbol{s}=[s_1 \ \ldots \ s_L]$ is a vector of $L$ different elements $s_l$, with each element $s_l$ relating to the magnitude of elements within $v$ and each $W_l$. Further,
\begin{align*}
a_{\boldsymbol{s}} &= \lambda+\left(\bar{\lambda}\tau\sum_{l \in [L]}s_l\right)-
    \left(\tau\sum_{l \in [L]}\sum_{j \in [s_l]}|U_{l,(j)}|\right).
\end{align*}
\end{theorem}

\begin{proof} We break the proof into two parts: first by solving for each $W^*_l$, and then solving for $b^*$. \\

\noindent \textbf{\textit{Derivation for $\boldsymbol{W^*_l}$.}}  \\

To derive closed-form expressions for each $W^{*}_l$, $l=1,\ldots,L$, we first simplify the optimization problem from (\ref{eq:optim}) to
\begin{align*}
    \left\{W^*_{1},\ldots,W^*_{L}\right\} &= \argmin_{W_{1},\ldots,W_{L}} \ \ \ \frac{1}{2} \sum_{l \in [L]}||U_l-W_l||^2_2 +\sum_{l \in [L]}\bar{\lambda}||W_l||_1 \\
    &\ \ \ \mathrm{subject \ to}   \quad   ||W_l||_{\infty}\leq \tau||b^*||_2 \ \ \forall l\in [L],
\end{align*}
\noindent which excludes the terms that concern estimation of $b^*$ in the objective function. Further, each $W^*_{i}$ does not rely on $W^*_{j}$ for all $i\neq j$ when estimating $W^*_i$. As a consequence, a closed-form expression for each $W^*_l$ can be obtained separately, and the optimization problem for estimating each $W^*_l$ is given by
\begin{equation}
\begin{aligned}
    W^*_{l} &= \argmin_{W_{l}} \ \ \ \frac{1}{2} ||U_l-W_l||^2_2 +\bar{\lambda}||W_l||_1 \\
    &\ \ \ \mathrm{subject \ to}   \quad   ||W_{l}||_{\infty}\leq \tau||b^*||_2 .
\end{aligned} \label{eq:optim3}
\end{equation}
To solve the optimization problem in (\ref{eq:optim3}), first note that both the objective function and its constraint are convex, as well as that feasible solutions exist for this optimization problem. These conditions of convexity and feasibility indicate that Slater's condition holds, which implies that strong duality also holds. Strong duality indicates that there exists a dual variable $\nu_l\in \mathbb{R}^K_{+}$ such that $W_l^*$ will minimize the Lagrangian
\begin{align*}
    W^{*}_l&=\argmin_{W_l} \ \ \ \frac{1}{2} ||U_l-W_l||^2_2 +\bar{\lambda}||W_l||_1+\sum_{j \in [K]}\nu_{l,j}(|W_{l,j}|-\tau||b^*||_2).
\end{align*}
\noindent Strong duality also indicates that each element of $W_l^*$, $W_{l,j}^*$ for $j = 1,\ldots,K$, must satisfy the following KKT conditions \citep[Section 5.5.3]{Boyd2004}:
\begin{enumerate}
    \item $W^{*}_{l,j}-U_{l,j}+(\bar{\lambda}+\nu_{l,j})\eta_{l,j}^*=0$ for some $\eta_{l,j}^*\in \delta(|W^{*}_{l,j}|)$ (stationarity).
    \item $\nu_{l,j}(|W^{*}_{l,j}|-\tau||b^*||_2)=0$ (complementary slackness).
    \item $\nu_{l,j}\geq 0$ (dual feasibility).
    \item $|W^{*}_{l,j}|\leq \tau||b^*||_2$ (primal feasibility).
\end{enumerate}
\noindent It is noted that for the stationarity condition, $\delta(|\cdot|)$ denotes the subderivative for the absolute value function, where 
\begin{equation}
\eta^*_{l,j} \in  \{-1\} \ \text{when $W_{l,j}^* < 0$}, \quad 
\eta^*_{l,j} \in  [-1, 1] \ \text{when $W_{l,j}^* = 0$}, \quad \eta^*_{l,j} \in  \{1\}  \ \text{when $W_{l,j}^* > 0$}. \label{eq:subdef}
\end{equation}

We first consider the case where $\nu_{l,j} = 0$ for the dual feasibility condition. In this case, the stationarity condition simplifies to
\begin{align*}
W^*_{l,j} - U_{l,j}+\bar{\lambda}\eta_{l,j}^*=0 
\end{align*}
for some $\eta_{l,j}^*\in \delta(|W_{l,j}^*|)$. In the scenario where $W_{l,j}^*\neq 0$, we would then obtain 
\begin{align}
W^*_{l,j} = U_{l,j}-\bar{\lambda}\eta_{l,j}^*. \label{eq:scen_w1}
\end{align} 
We note that $W^*_{l,j}$ and $\eta_{l,j}^*$ must also have the same sign by definition of the subderivative in (\ref{eq:subdef}), and as a consequence $U_{l,j}$ has the same sign as $W^*_{l,j}$ for the stationarity condition to hold. In the scenario where we instead have $W^*_{l,j}=0$, the stationarity condition now implies that 
\begin{align}
    U_{l,j} \in [-\bar{\lambda}, \bar{\lambda}]. \label{eq:scen_w2}
\end{align}
In other words, the stationarity condition would indicate that $W^*_{l,j}=0$ whenever $|U_{l,j}| \leq \bar{\lambda}$. Consideration of (\ref{eq:scen_w1}) when $W^*_{l,j}\neq0$ and (\ref{eq:scen_w2}) when $W^*_{l,j}=0$ yields
\begin{align}
    W^*_{l,j} &= \mathcal{S}_{\bar{\lambda}}(U_{l,j}), \label{eq:w1}
\end{align}
where $\mathcal{S}_{\bar{\lambda}}(U_{l,j})$ is the soft-thresholding operator from Definition \ref{def1}.
It is emphasized that when $\nu_{l,j}=0$, the primal feasibility condition implies that $|W^*_{l,j}|=\mathcal{S}_{\bar{\lambda}}(|U_{l,j}|)\leq \tau||b^*||_2$.

Now, consider the case where $\nu_{l,j}> 0$. In order to satisfy the complementary slackness condition, the equality $|W^*_{l,j}|=\tau||b^*||_2$ must hold. Further, from the stationarity condition, the equality
\begin{align*}
W^*_{l,j} = U_{l,j} - \left(\bar{\lambda}+\nu_{l,j}\right)\eta^*_{l,j}
\end{align*}
must also hold for some $\eta_{l,j}^*\in \delta(|W_{l,j}^*|)$. As $\mathrm{sign}(W_{l,j}^*)=\mathrm{sign}(\eta_{l,j}^*)$ by definition of the subderivative in (\ref{eq:subdef}) and  $\bar{\lambda}$ as well as $\nu_{l,j}$ are non-negative quantities, $\mathrm{sign}(W_{l,j}^*)=\mathrm{sign}(U_{l,j})$ for the stationarity condition to hold. We must conclude that
\begin{align}
    W_{l,j}^* = \mathrm{\mathrm{sign}}(U_{l,j})\tau||b^*||_2 \label{eq:ws2}
\end{align}
when $\nu_{l, j} > 0$. Combining (\ref{eq:w1}) and (\ref{eq:ws2}) together, we obtain
\begin{align}
    W_{l,j}^* &=\mathrm{sign}(U_{l,j})\mathrm{min}\left\{\tau||b^*||_2, \mathcal{S}_{\bar{\lambda}}(|U_{l,j}|)\right\}. \label{eq:w_sol}
\end{align}
\\
\noindent \textit{\textbf{Derivation for $\boldsymbol{b^*}$.}} \\

\noindent To derive the update for $b^*$, we note
\begin{align}
    \tau||b^*||_2 \in \left[\mathcal{S}_{\bar{\lambda}}(|U_{l,(s_l+1)}|), \mathcal{S}_{\bar{\lambda}}(|U_{l, (s_l)}|) \right) \label{eq:intervals}
\end{align}
for all $l=1,\ldots,L$. The identity in (\ref{eq:intervals}) is due to (\ref{eq:w_sol}), as $\mathcal{S}_{\bar{\lambda}}(|U_{l, (j)}|) \geq \tau||b||_2$ for all $j\leq s_l$. It is also noted that in this portion of the derivation we will express each $W^*_{l}$ as $W^*_{l}(b)$ given that (\ref{eq:w_sol}) depends on $b^*$.

Following Definition \ref{def2}, let $W^*_{l,(j)}(b)$ be the $j$-th sorted coordinate of $W^*_l(b)$, where $W^*_{l,(1)}(b)\geq W^*_{l,(2)}(b)\geq...\geq W^*_{l,(K)}(b)$. Further, given the optimization problem in (\ref{eq:optim}), define the objective function with respect to $b$ as
\begin{align}
    F(b) &= \frac{1}{2}\left(||v-b||_2^2+\sum_{l \in [L]}||U_{l}-W^*_{l}(b)||^2_2 \right)+\lambda||b||_2+\bar{\lambda}\sum_{l \in [L]}||W^*_{l}(b)||_1 \nonumber \\
    &= \frac{1}{2}\left[||v-b||_2^2+\sum_{l \in [L]}\sum_{j \in [s_l]}\left\{U_{l,(j)}-\mathrm{sign}(U_{l,(j)})\tau||b||_2\right\}^2 \right] \nonumber \\
    & \quad \quad +\lambda||b||_2+\bar{\lambda}\left\{\sum_{l \in [L]}\sum_{j \in [s_l]}\left|\mathrm{sign}(U_{l,(j)})\tau||b||_2\right|\right\}+c \label{r1} \\
    &= \frac{1}{2}\left[||v-b||_2^2+
    \sum_{l \in [L]}\sum_{j \in [s_l]}\left\{U_{l,(j)}-\mathrm{sign}(U_{l,(j)})\tau||b||_2\right\}^2 \right] \nonumber \\
    &\quad \quad +\lambda||b||_2+\bar{\lambda}\tau||b||_2\left(\sum_{l \in [L]}s_l\right)+c. \label{r2}
\end{align}
Here, (\ref{r1}) defines
\begin{align*}
    c &= \sum_{l\in [L]}\sum_{j \in \{s_l+1,\ldots, L\}}\left(\frac{\left\{U_l-\mathcal{S}_{\bar{\lambda}}(U_{l,j})\right\}^2}{2} + \bar{\lambda}\left|\mathcal{S}_{\bar{\lambda}}(U_{l,j})\right|\right),
\end{align*}
as $W^*_{l,(j)}(b)=\mathcal{S}_{\bar{\lambda}}(U_{l,(j)})$ when $j \geq s_l+1$ in (\ref{eq:w_sol}); and (\ref{r2}) is due to
\begin{align*}
\bar{\lambda}\sum_{l \in [L]}\sum_{j \in [s_l]}\left|\mathrm{sign}(U_{l,(j)})\tau||b||_2\right| = \bar{\lambda}\tau||b||_2\left(\sum_{l \in [L]}s_l\right).
\end{align*}
It is noted that $c$ will encompass terms that do not affect to estimation of $b^*$ for the remainder of this derivation.

To further simplify this objective function, we find
\begin{align*}
\sum_{l \in [L]}\sum_{j \in [s_l]}\left\{U_{l,(j)}-\mathrm{sign}(U_{l,(j)})\tau||b||_2\right\}^2 &=\sum_{l \in [L]}\sum_{j \in [s_l]}\left\{U_{l,(j)}^2-2U_{l,(j)}\mathrm{sign}(U_{l,(j)})\tau||b||_2+\tau^2||b||^2_2\right\} \\
&= \left\{\sum_{l \in [L]}\sum_{j \in [s_l]}\left(-2|U_{l,(j)}|\tau||b||_2+\tau^2||b||^2_2\right) \right\}+c \\
&= -2\tau||b||_2
    \left(\sum_{l \in [L]}\sum_{j \in [s_l]}|U_{l,(j)}|\right)+\tau^2||b||^2_2\left(\sum_{l \in [L]}s_l\right)+c,
\end{align*}
and it follows that
\begin{align}
    F(b) &= \frac{1}{2}\left\{||v-b||_2^2-2\tau||b||_2
    \left(\sum_{l \in [L]}\sum_{j \in [s_l]}|U_{l,(j)}|\right)+\tau^2||b||^2_2\left(\sum_{l \in [L]}s_l \right)\right\}\nonumber  \\
    &\quad \quad +\lambda||b||_2+\bar{\lambda}\tau||b||_2\left(\sum_{l \in [L]}s_l\right)+c. \nonumber 
\end{align}
Finally, $||v-b||_2^2 = ||b||_2^2-2v^Tb+||v||^2_2 = ||b||^2_2-2v^Tb+c$, which yields
\begin{align}
    F(b) &= \frac{1}{2}\left\{||b||_2^2-2v^Tb-2\tau||b||_2
    \left(\sum_{l \in [L]}\sum_{j \in [s_l]}|U_{l,(j)}|\right)+\tau^2||b||^2_2\left(\sum_{l \in [L]}s_l \right)\right\} \nonumber \\
    &\quad \quad +\lambda||b||_2+\bar{\lambda}\tau||b||_2\left(\sum_{l \in [L]}s_l\right)+c \nonumber \\ 
    &= \frac{1}{2}\left\{\left(1+\tau^2\sum_{l \in [L]}s_l\right)||b||_2^2-2v^Tb \right\}+\left\{\lambda+\bar{\lambda}\tau\left(\sum_{l \in [L]}s_l\right)-
    \tau\sum_{l \in [L]}\sum_{j \in [s_l]}|U_{l,(j)}|\right\}||b||_2+c \nonumber \\
    &= \frac{1}{2}\left\{\left(1+\tau^2\sum_{l \in [L]}s_l\right)||b||_2^2-2v^Tb \right\}+a_{\boldsymbol{s}}||b||_2+c, \nonumber
\end{align}
where
\begin{align*}
    a_{\boldsymbol{s}} &= \lambda+\bar{\lambda}\tau\left(\sum_{l \in [L]}s_l\right)-
    \tau\sum_{l \in [L]}\sum_{j \in [s_l]}|U_{l,(j)}|.
\end{align*}

Given that $F(b)$ is convex, $b^*$ can be obtained from the first-order optimality condition
\begin{align*}
    \nabla F(b^*) &= \left\{\left(1+\tau^2\sum_{l \in [L]}s_l\right)b^*-v \right\}+\frac{a_{\boldsymbol{s}}b^*}{||b^*||_2}=0
\end{align*}
when $b^*\neq 0$. Specifically,
\begin{align*}
\left\{\left(1+\tau^2\sum_{l \in [L]}s_l\right)+ \frac{a_{\boldsymbol{s}}}{||b^*||_2}\right\}b^*-v &= 0,
\end{align*}
which yields
\begin{align}
b^*&=\left\{\left(1+\tau^2\sum_{l \in [L]}s_l\right)+ \frac{a_{\boldsymbol{s}}}{||b^*||_2}\right\}^{-1}v. \label{eq:b_1}
\end{align}
Taking the $L_2$ norm on both sides of (\ref{eq:b_1}) results in
\begin{align*}
    ||b^*||_2 &= \left\{\left(1+\tau^2\sum_{l \in [L]}s_l\right)+\frac{a_{\boldsymbol{s}}}{||b^*||_2}\right\}^{-1}||v||_2,
\end{align*}
where this expression can be stated in terms of $||v||_2$ as
\begin{align*}
||v||_2 &=||b^*||_2\left\{\left(1+\tau^2\sum_{l \in [L]}s_l\right)+\frac{a_{\boldsymbol{s}}}{||b^*||_2}\right\} \\
    &=||b^*||_2\left(1+\tau^2\sum_{l \in [L]}s_l\right)+a_{\boldsymbol{s}}.
\end{align*}
This enable identification of 
\begin{align*}
    ||b^*||_2 &= \frac{||v||_2-a_{\boldsymbol{s}}}{(1+\tau^2\sum_{l \in [L]}s_l)},
\end{align*}
and we can conclude that $b^*$ equals
\begin{align}
    b^* &= \left\{\left(1+\tau^2\sum_{l \in [L]}s_l\right)+ \frac{a_{\boldsymbol{s}}}{\frac{||v||_2-a_{\boldsymbol{s}}}{(1+\tau^2\sum_{l \in [L]}s_l)}}\right\}^{-1}v \nonumber \\
    &= \left(\frac{1}{1+\tau^2\sum_{l \in [L]}s_l}\right)\left(1+\frac{a_{\boldsymbol{s}}}{||v||_2-a_{\boldsymbol{s}}} \right)^{-1}v \nonumber \\
    &= \left(\frac{1}{1+\tau^2\sum_{l \in [L]}s_l}\right)\left(1-\frac{a_{\boldsymbol{s}}}{||v||_2}\right)v \label{eq:bs_1}
\end{align}
when $b^*\neq 0$. 

If it is instead the case that $b^* = 0$, the subderivative of $F(b)$ yields
\begin{align}
    \left\{\left(1+\tau^2\sum_{l \in [L]}s_l\right)b^*-v \right\}+a_{\boldsymbol{s}}\eta^*=-v +a_{\boldsymbol{s}}\eta^*=0, \label{eq:almost}
\end{align}
where $\eta^* \in \delta(||b^*||_2) = \{\eta^* \ | \ ||\eta^*||_2 \leq 1\}$. The equality in (\ref{eq:almost}) implies that
\begin{align*}
    \frac{\left|\left|v\right|\right|_2}{a_{\boldsymbol{s}}} = ||\eta^*||_2^2 \leq 1,
\end{align*}
which indicates that $b^* = 0$ whenever $||v||_2 \leq a_{\boldsymbol{s}}$. Combining this and (\ref{eq:bs_1}) together, we conclude
\begin{align*}
b^* = \left(\frac{1}{1+\tau^2\sum_{l \in [L]}s_l}\right)\mathrm{max}\left(1-\frac{a_{\boldsymbol{s}}}{||v||_2}, 0\right)v.
\end{align*}
\end{proof}

\subsection{Search Procedure for $\boldsymbol{s}$} \label{as:app1.2}

The proximal step derived in Theorem \ref{thm1} depend on an additional parameter, $\boldsymbol{s} = [s_1 \ \ldots \ s_L]$. To determine $\boldsymbol{s}$, a search procedure is first applied to identify all candidate vectors $\boldsymbol{s}^{\dagger}$ that satisfy the condition
$$ \tau||b(\boldsymbol{s}^{\dagger})||_2 \in \left[ \mathcal{S}_{\bar{\lambda}}( |U_{l, (s^{\dagger}_l+1)}| ), \mathcal{S}_{\bar{\lambda}}( |U_{l, (s^{\dagger}_l)}|) \right) $$ 
for all $l \in [L]$ from (\ref{eq:intervals}). The parameter $\boldsymbol{s}$ is then defined as the vector that minimizes the optimization problem in (\ref{eq:optim}) among the identified candidates given the solutions $b^*$ and each $W^*_l$ derived in Theorem \ref{thm1} relative to each identified candidate $\boldsymbol{s}^{\dagger}$.

\bibliographystyle{apalike} 
\bibliography{bibliography}

\begin{thebibliography}{}

\bibitem[Archer et~al., 2014]{Archer2014}
Archer, S.~N., Laing, E.~E., M\"{o}ller-Levet, C.~S., van~der Veen, D.~R., Bucca, G., Lazar, A.~S., Santhi, N., Slak, A., Kabiljo, R., von Schantz, M., Smith, C.~P., and Dijk, D.-J. (2014).
\newblock Mistimed sleep disrupts circadian regulation of the human transcriptome.
\newblock {\em Proceedings of the National Academy of Sciences}, 111(6).

\bibitem[Beck, 2017]{Beck2017}
Beck, A. (2017).
\newblock {\em First-Order Methods in Optimization}.
\newblock Society for Industrial and Applied Mathematics.

\bibitem[Bergstra and Bengio, 2012]{Bergstra2012}
Bergstra, J. and Bengio, Y. (2012).
\newblock Random search for hyper-parameter optimization.
\newblock {\em Journal of Machine Learning Research}, 13:281–305.

\bibitem[Blatter and Cajochen, 2007]{Blatter2007}
Blatter, K. and Cajochen, C. (2007).
\newblock Circadian rhythms in cognitive performance: Methodological constraints, protocols, theoretical underpinnings.
\newblock {\em Physiology \& Behavior}, 90(2–3):196–208.

\bibitem[Boyd and Vandenberghe, 2004]{Boyd2004}
Boyd, S. and Vandenberghe, L. (2004).
\newblock {\em Convex Optimization}.
\newblock Cambridge University Press, Cambridge, England.

\bibitem[Braun et~al., 2018]{Braun2018}
Braun, R., Kath, W.~L., Iwanaszko, M., Kula-Eversole, E., Abbott, S.~M., Reid, K.~J., Zee, P.~C., and Allada, R. (2018).
\newblock Universal method for robust detection of circadian state from gene expression.
\newblock {\em Proceedings of the National Academy of Sciences}, 115(39).

\bibitem[Dallmann et~al., 2016]{Dallmann2016}
Dallmann, R., Okyar, A., and Lévi, F. (2016).
\newblock Dosing-time makes the poison: Circadian regulation and pharmacotherapy.
\newblock {\em Trends in Molecular Medicine}, 22(5):430–445.

\bibitem[Duffy et~al., 2011]{Duffy2011}
Duffy, J.~F., Cain, S.~W., Chang, A.-M., Phillips, A. J.~K., M\"{u}nch, M.~Y., Gronfier, C., Wyatt, J.~K., Dijk, D.-J., Wright, K.~P., and Czeisler, C.~A. (2011).
\newblock Sex difference in the near-24-hour intrinsic period of the human circadian timing system.
\newblock {\em Proceedings of the National Academy of Sciences}, 108(supplement{\_}3):15602--15608.

\bibitem[Geladi and Kowalski, 1986]{Geladi1986}
Geladi, P. and Kowalski, B.~R. (1986).
\newblock Partial least-squares regression: a tutorial.
\newblock {\em Analytica Chimica Acta}, 185:1–17.

\bibitem[Gers et~al., 2003]{Gers2003}
Gers, F.~A., Schraudolph, N.~N., and Schmidhuber, J. (2003).
\newblock Learning precise timing with lstm recurrent networks.
\newblock {\em J. Mach. Learn. Res.}, 3:115--143.

\bibitem[Goodfellow et~al., 2016]{Goodfellow2016}
Goodfellow, I., Bengio, Y., and Courville, A. (2016).
\newblock {\em Deep Learning}.
\newblock Adaptive Computation and Machine Learning series. MIT Press, London, England.

\bibitem[Gorczyca, 2024]{Gorczycac2024}
Gorczyca, M.~T. (2024).
\newblock A mixed effects cosinor modelling framework for circadian gene expression.
\newblock {\em arXiv}.

\bibitem[Gorczyca et~al., 2024a]{Gorczycaa2024}
Gorczyca, M.~T., McDonald, T.~M., and Sefas, J.~D. (2024a).
\newblock A corrected score function framework for modelling circadian gene expression.
\newblock {\em Stat}, 13(4).

\bibitem[Gorczyca et~al., 2024b]{Gorczycad2024}
Gorczyca, M.~T., McDonald, T.~M., and Sefas, J.~D. (2024b).
\newblock A note on trigonometric regression in the presence of berkson-type measurement error.
\newblock {\em Statistica Neerlandica}, 78(4).

\bibitem[Gorczyca and Sefas, 2024]{Gorczycab2024}
Gorczyca, M.~T. and Sefas, J.~D. (2024).
\newblock On weighted trigonometric regression for suboptimal designs in circadian biology studies.
\newblock {\em arXiv}.

\bibitem[Graves, 2012]{Graves2012}
Graves, A. (2012).
\newblock Sequence transduction with recurrent neural networks.
\newblock {\em CoRR}, abs/1211.3711.

\bibitem[Graves et~al., 2013]{Graves2013}
Graves, A., Mohamed, A.-r., and Hinton, G. (2013).
\newblock Speech recognition with deep recurrent neural networks.
\newblock {\em arXiv}.

\bibitem[Graves and Schmidhuber, 2005]{Graves2005}
Graves, A. and Schmidhuber, J. (2005).
\newblock Framewise phoneme classification with bidirectional lstm and other neural network architectures.
\newblock {\em Neural Networks}, 18(5–6):602–610.

\bibitem[Hastie et~al., 2009]{Hastie2009}
Hastie, T., Tibshirani, R., and Friedman, J. (2009).
\newblock {\em The Elements of Statistical Learning}.
\newblock Springer, New York, NY.

\bibitem[He et~al., 2016]{He2016}
He, K., Zhang, X., Ren, S., and Sun, J. (2016).
\newblock Deep residual learning for image recognition.
\newblock In {\em 2016 IEEE Conference on Computer Vision and Pattern Recognition (CVPR)}. IEEE.

\bibitem[Hochreiter and Schmidhuber, 1997]{Hochreiter1997}
Hochreiter, S. and Schmidhuber, J. (1997).
\newblock Long short-term memory.
\newblock {\em Neural Computation}, 9(8):1735–1780.

\bibitem[Huang and Braun, 2024]{Huang2024}
Huang, Y. and Braun, R. (2024).
\newblock Platform-independent estimation of human physiological time from single blood samples.
\newblock {\em Proceedings of the National Academy of Sciences}, 121(3).

\bibitem[Hughey, 2017]{Hughey2017}
Hughey, J.~J. (2017).
\newblock Machine learning identifies a compact gene set for monitoring the circadian clock in human blood.
\newblock {\em Genome Medicine}, 9(1).

\bibitem[Kantermann et~al., 2015]{Kantermann2015}
Kantermann, T., Sung, H., and Burgess, H.~J. (2015).
\newblock Comparing the morningness-eveningness questionnaire and munich {chronoType} questionnaire to the dim light melatonin onset.
\newblock {\em Journal of Biological Rhythms}, 30(5):449--453.

\bibitem[Kennaway, 2019]{Kennaway2019}
Kennaway, D.~J. (2019).
\newblock A critical review of melatonin assays: past and present.
\newblock {\em Journal of Pineal Research}, 67(1).

\bibitem[Kennaway, 2023]{Kennaway2023}
Kennaway, D.~J. (2023).
\newblock The dim light melatonin onset across ages, methodologies, and sex and its relationship with morningness/eveningness.
\newblock {\em {Sleep}}, 46(5).

\bibitem[Laing et~al., 2017]{Laing2017}
Laing, E.~E., M\"{o}ller-Levet, C.~S., Poh, N., Santhi, N., Archer, S.~N., and Dijk, D.-J. (2017).
\newblock Blood transcriptome based biomarkers for human circadian phase.
\newblock {\em eLife}, 6.

\bibitem[Lane et~al., 2022]{Lane2022}
Lane, J.~M., Qian, J., Mignot, E., Redline, S., Scheer, F. A. J.~L., and Saxena, R. (2022).
\newblock Genetics of circadian rhythms and sleep in human health and disease.
\newblock {\em Nature Reviews Genetics}, 24(1):4–20.

\bibitem[Lemhadri et~al., 2021]{Lemhadri2021}
Lemhadri, I., Ruan, F., Abraham, L., and Tibshirani, R. (2021).
\newblock Lassonet: A neural network with feature sparsity.
\newblock {\em Journal of Machine Learning Research}, 22(127):1--29.

\bibitem[Levi and Schibler, 2007]{Levi2007}
Levi, F. and Schibler, U. (2007).
\newblock Circadian rhythms: Mechanisms and therapeutic implications.
\newblock {\em Annual Review of Pharmacology and Toxicology}, 47(1):593–628.

\bibitem[Lewy, 1999]{Lewy1999}
Lewy, A.~J. (1999).
\newblock The dim light melatonin onset, melatonin assays and biological rhythm research in humans.
\newblock {\em Neurosignals}, 8(1-2):79--83.

\bibitem[Long et~al., 2016]{Long2016}
Long, J.~E., Drayson, M.~T., Taylor, A.~E., Toellner, K.~M., Lord, J.~M., and Phillips, A.~C. (2016).
\newblock Morning vaccination enhances antibody response over afternoon vaccination: A cluster-randomised trial.
\newblock {\em Vaccine}, 34(24):2679–2685.

\bibitem[Malpas and Purdie, 1990]{Malpas1990}
Malpas, S.~C. and Purdie, G.~L. (1990).
\newblock Circadian variation of heart rate variability.
\newblock {\em Cardiovascular Research}, 24(3):210–213.

\bibitem[M\"{o}ller-Levet et~al., 2013]{MllerLevet2013}
M\"{o}ller-Levet, C.~S., Archer, S.~N., Bucca, G., Laing, E.~E., Slak, A., Kabiljo, R., Lo, J. C.~Y., Santhi, N., von Schantz, M., Smith, C.~P., and Dijk, D.-J. (2013).
\newblock Effects of insufficient sleep on circadian rhythmicity and expression amplitude of the human blood transcriptome.
\newblock {\em Proceedings of the National Academy of Sciences}, 110(12).

\bibitem[Montaigne et~al., 2018]{Montaigne2018}
Montaigne, D., Marechal, X., Modine, T., Coisne, A., Mouton, S., Fayad, G., Ninni, S., Klein, C., Ortmans, S., Seunes, C., Potelle, C., Berthier, A., Gheeraert, C., Piveteau, C., Deprez, R., Eeckhoute, J., Duez, H., Lacroix, D., Deprez, B., Jegou, B., Koussa, M., Edme, J.-L., Lefebvre, P., and Staels, B. (2018).
\newblock Daytime variation of perioperative myocardial injury in cardiac surgery and its prevention by rev-erba antagonism: a single-centre propensity-matched cohort study and a randomised study.
\newblock {\em The Lancet}, 391(10115):59–69.

\bibitem[Paszke et~al., 2019]{pytorch2021}
Paszke, A., Gross, S., Massa, F., Lerer, A., Bradbury, J., Chanan, G., Killeen, T., Lin, Z., Gimelshein, N., and Antiga, L. (2019).
\newblock Pytorch: An imperative style, highperformance deep learning library.
\newblock In {\em Conference on Neural Information Processing Systems (NeurIPS)}.

\bibitem[{R Core Team}, 2021]{R2021}
{R Core Team} (2021).
\newblock {\em R: A Language and Environment for Statistical Computing}.
\newblock R Foundation for Statistical Computing, Vienna, Austria.

\bibitem[Refinetti and Menaker, 1992]{Refinetti1992}
Refinetti, R. and Menaker, M. (1992).
\newblock The circadian rhythm of body temperature.
\newblock {\em Physiology \& Behavior}, 51(3):613–637.

\bibitem[Reid, 2019]{Reid2019}
Reid, K.~J. (2019).
\newblock Assessment of circadian rhythms.
\newblock {\em Neurologic Clinics}, 37(3):505--526.

\bibitem[Ruben et~al., 2019]{Ruben2019}
Ruben, M.~D., Smith, D.~F., FitzGerald, G.~A., and Hogenesch, J.~B. (2019).
\newblock Dosing time matters.
\newblock {\em Science}, 365(6453):547–549.

\bibitem[Ruiz et~al., 2020]{Ruiz2020}
Ruiz, F.~S., Beijamini, F., Beale, A.~D., da~Silva B.~Gon{\c{c}}alves, B., Vartanian, D., Taporoski, T.~P., Middleton, B., Krieger, J.~E., Vallada, H., Arendt, J., Pereira, A.~C., Knutson, K.~L., Pedrazzoli, M., and von Schantz, M. (2020).
\newblock Early chronotype with advanced activity rhythms and dim light melatonin onset in a rural population.
\newblock {\em Journal of Pineal Research}, 69(3).

\bibitem[Schuster and Paliwal, 1997]{Schuster1997}
Schuster, M. and Paliwal, K. (1997).
\newblock Bidirectional recurrent neural networks.
\newblock {\em IEEE Transactions on Signal Processing}, 45(11):2673–2681.

\bibitem[Tibshirani, 1996]{Tibshirani1996}
Tibshirani, R. (1996).
\newblock Regression shrinkage and selection via the lasso.
\newblock {\em Journal of the Royal Statistical Society Series B: Statistical Methodology}, 58(1):267–288.

\bibitem[Wittenbrink et~al., 2018]{Wittenbrink2018}
Wittenbrink, N., Ananthasubramaniam, B., M\"{u}nch, M., Koller, B., Maier, B., Weschke, C., Bes, F., de~Zeeuw, J., Nowozin, C., Wahnschaffe, A., Wisniewski, S., Zaleska, M., Bartok, O., Ashwal-Fluss, R., Lammert, H., Herzel, H., Hummel, M., Kadener, S., Kunz, D., and Kramer, A. (2018).
\newblock High-accuracy determination of internal circadian time from a single blood sample.
\newblock {\em Journal of Clinical Investigation}, 128(9):3826--3839.

\bibitem[{World Health Organization}, 2010]{WHO2010}
{World Health Organization} (2010).
\newblock {\em WHO Guidelines on Drawing Blood: Best Practices in Phlebotomy}.
\newblock World Health Organization Press, Geneva, Switzerland.

\bibitem[Wright et~al., 2013]{Wright2013}
Wright, K.~P., McHill, A.~W., Birks, B.~R., Griffin, B.~R., Rusterholz, T., and Chinoy, E.~D. (2013).
\newblock Entrainment of the human circadian clock to the natural light-dark cycle.
\newblock {\em Current Biology}, 23(16):1554--1558.

\end{thebibliography}

\end{document}